\documentstyle[12pt,procsla]{article}
\catcode`\@=11
\long\def\@makefntext#1{
\protect\noindent \hbox to 3.2pt {\hskip-.9pt
$^{{\ninerm\@thefnmark}}$\hfil}#1\hfill}                

\def\@makefnmark{\hbox to 0pt{$^{\@thefnmark}$\hss}}  

\def\ps@myheadings{\let\@mkboth\@gobbletwo
\def\@oddhead{\hbox{}
\rightmark\hfil\ninerm\thepage}
\def\@oddfoot{}\def\@evenhead{\ninerm\thepage\hfil
\leftmark\hbox{}}\def\@evenfoot{}
\def\sectionmark##1{}\def\subsectionmark##1{}}

\begin{document}

\title{
\bf UNIVERSAL MASS DEPENDENCE
FOR PARTICLE PRODUCTION RATES
IN $e^+e^-$, $pp$ AND $AA$ COLLISIONS
FROM THE QUARK-GLUON PLASMA PERSPECTIVE}
\author{M.~Szczekowski and
G.~Wilk\\
\centerline{\it Soltan Institute for Nuclear Studies,
Ho\.za 69, PL-00-681 Warsaw, Poland}}
\date{}
\maketitle
\vspace{6cm}
Talk presented at {\it XXV International
Symposium on Multiparticle Dynamics}, Star\'a Lesn\'a,
High Tatra Mountains, SLOVAKIA, 12-16 September 1995;
to be published in the Proceedings (eds. D.Bruncko, L.Sandor and
J.Urban).\\

\vspace{2cm}
\centerline{SINS preprint: PVI and PVIII/1995-25 }

\newpage

\centerline{\normalsize\bf UNIVERSAL MASS DEPENDENCE}
\baselineskip=21pt
\centerline{\normalsize\bf FOR PARTICLE PRODUCTION RATES}
\baselineskip=21pt
\centerline{\normalsize\bf IN $e^+e^-$, $pp$ AND $AA$ COLLISIONS}
\baselineskip=21pt
\centerline{\normalsize\bf FROM THE QUARK-GLUON PLASMA PERSPECTIVE}
\baselineskip=21pt
\centerline{\footnotesize M.~Szczekowski$^{\dag}$ and
\underline{G.~Wilk}$^{\ddag}$}
\baselineskip=13pt
\centerline{\footnotesize\it Soltan Institute for Nuclear Studies,
Ho\.za 69, PL-00-681 Warsaw, Poland}
\baselineskip=12pt
\centerline{\footnotesize E-mail: $^{\dag}$szczekow@cernvm.cern.ch;~~
                                  $^{\ddag}$wilk@fuw.edu.pl}
\vspace*{0.9cm}
\abstracts{It is demonstrated that universal mass dependence for
meson and baryon inclusive cross-sections advocated recently in
$e^+e^-$ annihilation and in hadronic reactions is consistent also
with heavy ion collisions data and can therefore be used as a
reference for quark-gluon plasma studies.}

\normalsize\baselineskip=15pt
\setcounter{footnote}{0}
\renewcommand{\thefootnote}{\alph{footnote}}

\section{Introduction}
The search for the new state of matter, the quark-gluon plasma (QGP),
is nowadays one of the main driving forces behind the constant
interest in multiparticle production processes at high energies.
There is a vast literature discussing theoretical and experimental
aspects of possible signals of QGP \cite{QM}. One of the possible
signals of the QGP is the enhancement of strangeness production
in heavy ion collisions. Interpretation of this enhancement can be,
however, regarded as still a controversial one with several models
proposed for the explanation of existing data which fall, roughly
speaking, into two categories:
\begin{itemize}
\item thermal models, including both those with QGP notion and models
exploring the hadron gas concept \cite{SH,LTHSR,PS};
\item models using, in one or another form, the concept of colour
string formation and its subsequent fragmentation
\cite{S,W,ABP,C}$^,$\footnote{There are attempts to provide a model
{\it  interpolating}, in a sense, between both alternatives presented
here, by starting with strings and thermalizing them afterwards in
such a way that particles are finally produced from a kind of thermal
fireball, cf. Ref. \cite{W}.}~.
\end{itemize}

We shall present here an observation which in our opinion can serve
as an additional check for possible models of multiparticle
production in heavy ion collisons, in particular to discriminate
between the two classes of models mentioned above with the ultimate
aim to help in clarifying further the problem of existence of the
QGP.

\section{Universal mass dependence in $e^+e^-$ annihilations and
hadronic reactions}
That observation is the apparent {\it universal mass dependence for
meson and baryon inclusive cross-sections} first observed in $e^+e^-$
annihilations \cite{CU} and recently confirmed also in hadronic
reactions \cite{SZ}. It was demonstrated there that data on inclusive
particle production cross-sections (or rates of particles) in high
energy interactions follow the very simple formula:
\begin{equation}
\sigma_{incl} = a\, \frac{2J + 1}{2I_m + 1}\,
                \exp \left( -b M^2 \right)  \label{eq:F}
\end{equation}
where $J$ denotes the spin, $I_m$ a modified isospin and $M$ the mass
of the produced particle. Two points should be stressed here:
\begin{itemize}
\item the cross section depends on the mass of produced particle {\it
quadratically} and
\item it contains the spin ($2J+1$) and (modified) isospin
($2I_m+1$) weighting factors accounting roughly speaking for
non-observed states \footnote{Cf. refs. \cite{CU,SZ} for detailed
explanation of these weighting factors.}~.
\end{itemize}
In Fig. 1 (taken from \cite{CU}) one can see very clearly
that, except of pions, all scalar and vector mesons put together with
octet and decuplet baryons for the LEP energy lie, when plotted as a
function of $M^2$, on one  straight line with a slope parameter
\begin{equation}
b~=~3.9~{\rm (GeV/c}^2{\rm )}^{-2}. \label{eq:slope}
\end{equation}
The important thing here is that:
\begin{itemize}
\item the same value of the slope parameter $b$ is observed at both
PEP/PETRA and at LEP energies and that
\item only the normalization constant $a$ is energy dependent.
\end{itemize}
As one can see in Fig. 1:
\begin{itemize}
\item pions evidently do not follow this simple rule (most probably
as a result of the contribution of pions from the decays of high mass
resonances, but there are also other more speculative possibilities
mentioned by the authors of \cite{CU});
\item there is a deviation from this universal curve observed for the
heaviest particles produced at lower, i.e., PEP/PETRA, energy (cf.,
however, discussion in \cite{CU}~); this fact will be important
in our further discussion below.
\end{itemize}
The word {\it universal} used above refers to the fact that (almost)
all species of produced particles lie on the universal curve with the
same slope. But so far the figures presented here refer to only one
type of reactions, although the most elementary one: $e^+e^-
\rightarrow hadrons$. A follow-up question rises then immediately:
{\it Does this universality hold also for hadronic collisions?}\\

There are, however, two difficulties one has to be aware of
concerning hadronic collisions:
\begin{itemize}
\item[$(i)$] the available data are scarce and with large systematic
errors due to the well known experimental difficulties in the
identification of particles and  in the measurement of inclusive
cross-sections for hadronic resonances in a many particle
environment;
\item[$(ii)$] there is a difference between $e^+e^-$ annihilations
and $pp$ collisions: in hadronic reactions the quantum numbers of the
initial state are different from those of the vacuum state. It leads
to initial asymmetry which, for instance, in $pp$ interactions
manifests itself in the enhancement of the production of baryon
resonances in comparison to antibaryons because there are already two
baryons present in the initial state \footnote{In fact, one can argue
that there is also the third difficulty expressed by the fact that,
as far as strings are concerned, their configurations are more
complicated in hadronic collisions than in the $e^+e^-$
annihilations. We can expect, however, that some features of the
local hadron production will be similar in both types of
processes.}~.
\end{itemize}
The following sets of data were considered in \cite{SZ} and are shown
in Fig. 2 (taken from \cite{SZ}): $(i)$ - the analysis of $pp$
collisions done by LEBC-EHC Collab. for $p_{LAB} = 400$ GeV/c
(corresponding to center of mass energy $\sqrt{s} = 27.4$ GeV
\cite{A-B}), $(ii)$ - Fermilab 30-inch bubble chamber data at
$p_{LAB} = 405$ GeV/c ($\sqrt{s} = 27.6$ GeV) \cite{K} and $(iii)$ -
results from two ISR experiments: the SFM and ACCHMN Collab. at
$\sqrt{s} = 53$ GeV \cite{D}$^,$\footnote{The results from the two
ISR experiments for the production of high mass resonances are not
consistent and for $K^{\star}(1430)$ they differ by as much as an
order of magnitude!}~. Only the results for antibaryons and
neutral or negative mesons were considered because these particles
are produced mainly from the quark-antiquark or diquark-antidiquark
pairs created during string fragmentation and therefore should not be
sensitive to the possible initial state biases. The inclusive
cross-section for antiprotons at $\sqrt{s} = 53$ GeV was calculated
from the compilation \cite{R}.\\

Fig. 2 shows that the answer to the question posed above is positive.
For low mass particles we indeed observe the same regularity in meson
and baryon production also in high energy hadron-hadron collisions
and the corrected inclusive cross-sections are consistent with the
behaviour decribed by eq.(\ref{eq:F}) with the slope parameter $b$
taken from $e^+e^-$ annihilation data.\\

The observation of {\it universal} (i.e., $M^2$) behaviour of the
particle production rates points strongly towards the common
fragmentation/production mechanism operating in both types of
collisions, namely to hadron states being predominantly produced
locally in string fragmentation irrespective of the configuration of
the strings in various processes. To further check this conjecture
one would need, however, precise data on meson and (anti)baryon
inclusive cross-sections both for $\bar{p}p$ CERN Collider and
Fermilab Tevatron. Only then strings produced in hadronic collisions
will have masses comparable to those seen in $e^+e^-$ annihilations
at LEP. Hadronic data presented in Fig. 2 correspond in
this respect to $e^+e^-$ annihilations at the PEP/PETRA energies
where, as we have already noticed before, an upward deviation from
the universality has been  observed. We can attribute it to the final
energy effect originating from the non-universal (i.e., flavour
dependent) fragmentation of the ends of short strings. In hadronic
collisions we have also contributions from diffractive excitations.\\

\section{Universal mass dependence for particle production rates from
the Quark-Gluon Plasma perspective}
Encouraged by the analysis presented so far one can finally ask {\it
whether the similar universality can be found also in the most
complex multiparticle production processes, namely heavy ion
collisions?} \\

Such universal dependence, {\it quadratic} in the masses of produced
secondaries, points strongly towards particles being produced locally
in colour string fragmentation \cite{CB} rather then from any kind of
locally equilibrated (thermally and/or chemically) clusters or
fireballs which would lead to $\exp ( - \beta M ) $ or $\exp ( -
\beta \langle M_T\rangle ) $ factors instead (with $\beta$ being an
inverse temperature of the local thermal equilibrium and $\langle
M_T\rangle $ a mean transverse mass) \cite{H}. In this case from the
tunneling phenomenon $b=\pi/\kappa$ and can be interpreted as a
measure of the colour field strength $\kappa$ in strings.\\

As far as the QGP is concerned, its production must {\it by
definition} be connected with some sort of local thermal and chemical
equilibrium taking place on the level of quarks and gluons.
Therefore, from the point of view of the analysis presented here, in
high energy heavy ion collisions one should observe a departure from
the universal behaviour given by eq.(\ref{eq:F}) towards the formula
with the linear mass dependence expected in thermal models. This is
the main point of this presentation. We shall therefore check the
existing data on particle production in  heavy ion collisions for
this possibility. We can hope to differentiate in this way between
the two types of models mentioned at the beginning, only one of which
admit the presence of the QGP.\\

In Fig. 3 we present the available results for particle production
rates with $(2I_m+1)/(2J+1)$ weights accounted for in, respectively,
$S+S$ and $S+Ag$ collisions at $200$ GeV/nucleon and compare them
with the dependence given by eq.(\ref{eq:F}) with the slope parameter
(eq.(\ref{eq:slope})) taken from the $e^+e^-$ annihilation data fit
and normalized to the weighted kaon production rates. The data points
for $\pi^-$, $K^-$ and $\bar{\Lambda}$ production rates for central
collisions in full phase space are from \cite{DATA1}. The data for
$\bar{p}$ are calculated from  the ratios $\bar{\Lambda}/\bar{p} =
1.5\pm0.5$ (for $S+S$ collisions) and $0.8\pm0.25$ (for the $S+Ag$)
obtained from the fit to the corresponding central rapidity regions
\cite{DATA3}.\\

Notice that we again face here the same difficulty as in the case of
hadronic collisions: because, contrary to the $e^+e^-$ annihilation
case, quantum numbers of the initial state are different from those
of the vacuum state, we have to use only the measured rates for
antibaryons and neutral or negative mesons. It can be seen that,
again apart from pions, the universality of particle production
observed in $e^+e^-$ and in hadronic reactions holds also here (up to
the production of $\bar{\Lambda}$, which is the highest mass state
available at present for such analysis \footnote{Data for $\bar{\Xi}$
are obtained only in the higher transverse momentum region $1.0 < p_T
< 2.5$ GeV/c \cite{DATA2} and those for $\bar{\Omega}$ production
available so far are in the form unsuitable for the analysis
presented here \cite{DATA3}.}~). The slight deviation from the
universal curve for $\bar{\Lambda}$ can be observed only for heavier
nuclear targets. \\

Taking therefore the dependence (\ref{eq:F}) as the reference for
particle production rates we can conclude that the production of
$\bar{\Lambda}$ does not point toward the necessity of introduction
of the thermal models (where QGP can be possibly found). Data for
heavier strange antibaryons will be crucial here as they will tell us
whether the observed universality in $M^2$ extends also to higher
mass states or we shall see the first signs of an upward deviation.
However, in order for such data to be decisive we have to wait for
much higher energies then those available today. The point is that
any upward deviation which looks like an outset of "thermal" (i.e.,
linear in mass $M$) dependence can be at present energies linked in a
natural way to  the deviations seen already in $e^+e^-$ annihilation
at lower energies (cf. Fig. 1) and in proton-proton
scattering (cf. Fig. 2). One could then argue that at
present energies only strings of relatively low masses are formed in
the elementary processes constituting a nuclear collision. This in
turn would result in the upward departure from the universality for
high mass hadrons due to the predominant contributions from the
non-universal (i.e., flavour dependent) fragmentation of the ends of
strings taking over its universal breaking.\\

This explanation corresponds to the (again non thermal and surely non
QGP) mechanism proposed recently in \cite{C}. Its characteristic
feature is that, according to results of \cite{CU}, it should
decrease with increasing energies. Also string models operating with
colour rope dynamics  \cite{S}, what results in smaller
$b\sim1/\kappa$, should be able to cope with such deviation from the
universality. The same is true for new versions of {\it VENUS}
\cite{W} and {\it DTU} \cite{ABP} models using string fusion
concepts. This is a way of introduction of some collective production
mechanisms without any reference to QGP. Therefore experiments at
higher energies measuring very  precisely yields of all secondaries
should tell us which proposition is the correct one.\\

The weighted values for the pion production rates lie much higher
then predicted. The same effect is observed in $e^+e^-$ annihilations
and in proton-proton collisions. A compelling explanation for this
effect is the contribution of pions from the decays of high mass
resonances in addition to the direct pion production. From the QGP
point of view mesons, in particular pions, provide the bulk of the
observed entropy \cite{MG}. The similar behaviour of pion production
rates observed in all reactions considered here poses therefore an
additional challenge to the specifically QGP-type (or thermal)
oriented explanations for the copious pion production in heavy ion
collisions \cite{SHU}. \\

In conclusion:
\begin{itemize}
\item We have shown that the observed universality in the mass
dependence of the particle production rates can serve as an
additional useful tool for differentiating between multiparticle
production mechanisms. As in the hadronic case, the higher energy
data for heavy ion collisions will be of utmost interest here.
\item If the observed universality continues and extends to even
heavier particles (as it is observed in $e^+e^-$ annihilations at
LEP energies) it will be difficult for thermal models to account for
such behaviour \footnote{There are attempts to fit the particle rates
in $e^+e^-$ annihilation processes using a thermal model, cf., for
example Ref.\cite{BEC}. However, it does not provide such universal
behaviour as the one presented here and was not tested in all three
types of collisions.}~.
\item If, on the other hand, with growing energy and/or masses of
colliding nuclei this universality will break down even for lighter
particles, it would mean that some collective effects in the
production processes start to dominate; in this case thermal models
(containing possibly also QGP) will be the most effective ones in
description of this phenomenon.
\end{itemize}

\noindent
{\bf Acknowledgements:} This work was supported in part by the Polish
State Committee for Scientific Research grants 2P03B00108 and
SPUB/P03/103/95.

{\bf Figure Captions:}
\begin{itemize}
\item[{\bf Fig. 1}] Production rates of pseudoscalar and vector
mesons and octet and decuplet baryons at LEP $(a)$ and PETRA/PEP
energies $(b)$, weighted with the $(2I_m+1)/(2J+1)$ factor, as a
function of particle mass squared. The line shows the result of the
fit to the eq.(\ref{eq:F}).
\item[{\bf Fig. 2}] Inclusive cross-sections of mesons and
antibaryons for $pp$ collisions at $\sqrt{s}=27.4$ GeV $(a)$ and $53$
GeV $(b)$, weighted with the $(2I_m+1)/(2J+1)$ factor, as a function
of particle mass squared. The lines in both figures show the
eq.(\ref{eq:F}) dependence with the slope parameter the same as in
Fig. 1 and given by eq.(\ref{eq:slope}).
\item[{\bf Fig. 3}] Particle rates for negative and neutral mesons
and antibaryons produced in central $S+S$ collisions $(a)$ and in
central $S+Ag$ collisions $(b)$ (both at the energy $200$
GeV/nucleon) as a function of particle mass squared compared with the
fit from the $e^+e^-$ annihilation process.
\end{itemize}
\end{document}